\documentclass[aps,prd,showpacs,floatfix,preprint]{revtex4}
\usepackage{amsmath,bm}
\usepackage{graphicx}
\usepackage{epsfig}
\begin{document}

\title{Shear viscosity in antikaon condensed matter}

\author{ Rana Nandi$^1$, Sarmistha Banik$^2$ and Debades Bandyopadhyay$^1$}
\affiliation{$^1$Theory Division and Centre for Astroparticle Physics,
Saha Institute of Nuclear Physics, 1/AF Bidhannagar, 
Kolkata-700064, India}
\affiliation{$^2$Variable Energy Cyclotron Centre,
1/AF Bidhannagar, Kolkata-700064, India}

\begin{abstract}
We investigate the shear viscosity of neutron star matter in the presence of an
antikaon condensate. The electron and muon number densities are reduced 
due to the appearance of a $K^-$ condensate in neutron star matter, whereas 
the proton number density increases. Consequently the shear viscosity due to 
scatterings of electrons and muons with themselves and protons is lowered 
compared to the case without the condensate. On 
the other hand, the contribution of proton-proton collisions to the proton 
shear viscosity through electromagnetic and strong interactions, becomes 
important and comparable to the neutron shear viscosity.   
\end{abstract}

\maketitle

\section{Introduction}
Shear viscosity plays important roles in neutron star physics. It might damp 
the r-mode instability below the temperature $\sim 10^8$ K \cite{And}. The
knowledge of shear viscosity is essential in understanding pulsar glitches
and free precession of neutron stars \cite{Glam}. The calculation of the neutron
shear viscosity ($\eta_n$) for nonsuperfluid matter using free-space 
nucleon-nucleon scattering data was first done by Flowers and Itoh 
\cite{Flo1,Flo2}. Cutler and Lindblom \cite{Cut} fitted the results of Flowers
and Itoh \cite{Flo2} for the study of viscous damping of oscillations in 
neutron stars. Recently the neutron shear viscosity of pure neutron matter has 
been investigated in a self-consistent way \cite{Ben}.

It was noted that electrons, the lightest charged particles and neutrons, the 
most abundant particles in neutron star matter contribute significantly to 
the total shear viscosity. Flowers and Itoh
found that the neutron viscosity was larger than the combined viscosity of 
electrons and
muons ($\eta_{e\mu}$) in non-superfluid matter \cite{Flo2}. Further Cutler and
Lindblom argued that the electron viscosity was larger than the neutron 
viscosity in a superfluid neutron star \cite{Cut}. Later Andersson and his 
collaborators as well as Yakovlev and his collaborator showed 
$\eta_{e\mu} > \eta_n$ in the presence of proton superfluidity \cite{Glam,Yak}.
In the latter calculation, the effects of the exchange of transverse plasmons 
in the collisions of charged particles were included and it lowered the 
$\eta_{e\mu}$ compared with the case when only longitudinal plasmons were 
considered \cite{Yak}.   

So far, all of those calculations of shear viscosity were done in neutron star
matter composed of neutrons, protons, electrons and muons. 
However, exotic forms of matter such as hyperon or antikaon condensed matter
might appear in the interior of neutron stars. Negatively charged hyperons or 
a $K^-$ condensate could affect the electron shear viscosity appreciably. 

Here we focus on the role of $K^-$ meson condensates on the shear viscosity. 
No calculation of shear viscosity involving antikaon 
condensation has been carried out so far. This motivates us to investigate
the shear viscosity in the presence of an antikaon condensate. The
$K^-$ condensate appears at 2-3 times the normal nuclear matter density. With
the onset of the condensate, $K^-$ mesons replace electrons and muons in the
core. As a result, $K^-$ mesons along with protons maintain the charge 
neutrality. It was noted that the proton fraction became comparable to the 
neutron fraction in a neutron star including the $K^-$ condensate at higher 
densities \cite{Brow,Pal,Banik}. The
appearance of the $K^-$ condensate would not only influence the electron and
muon shear viscosities but it will also give rise to a new contribution called
the proton shear viscosity. 

This paper is organised in the following way. In Sec. II, we describe the 
calculation of shear viscosity in neutron stars involving the $K^-$ 
condensate. Results are discussed in Sec. III. A summary
is given in Sec. IV. 

\section{Formalism}

Here we are interested in calculating the shear viscosity of neutron star 
matter in the presence of an antikaon condensate.   
We consider neutron star matter undergoing a first order phase transition
from charge neutral and beta-equilibrated nuclear matter to a $K^-$ condensed 
phase. The nuclear phase is composed of neutrons, protons, electrons and muons
whereas the antikaon condensed phase is made up of neutrons and protons 
embedded in the Bose-Einstein condensate of $K^-$ mesons along with electrons 
and muons.
Antikaons form a s-wave ($\bf p =0$) condensation in this case. Therefore, 
$K^-$ mesons in the condensate do not take part in momentum transfer during
collisions with other particles. However, the condensate influences the proton
fraction and equation of state (EOS) which, in turn, might have important 
consequences for the shear viscosity. 
The starting point for the calculation of the shear viscosity is a set of
coupled Boltzmann transport equations \cite{Flo2,Yak} for the ith particle 
species  
(i= n, p, e, $\mu$) with velocity $v_i$ and distribution function $F_i$,

\begin{equation}
{\vec v_i} \cdot {\vec {\bigtriangledown}} F_i = {\sum}_{j=n,p,e,\mu} I_{ij}.~
\end{equation}
The transport equations are coupled through collision integrals given by,
\begin{equation} 
I_{ij} = {\frac{V^3}{(2\pi\hbar)^9 (1+\delta_{ij})}} 
{\sum_{{s_{i'}},{s_j},{s_{j'}}}}
\int d{\bf p}_jd{\bf p}_{i'}d{\bf p}_{j'} W_{ij}{\cal F}~,
\end{equation}
where
\begin{equation} 
{\cal F} = [{F_{i'}F_{j'} (1-F_{i})(1-F_{j})  - F_{i}F_{j} (1-F_{i'})
(1-F_{j'})}]~.  
\end{equation}
Here ${\bf p}_i$, ${\bf p}_j$ are momenta of incident particles and 
${\bf p}_{i'}$,
${\bf p}_{j'}$ are those of final states. The Kronecker delta in Eq. (2) is 
inserted to avoid double counting for identical particles. Spins are denoted by
$s$ and $W_{ij}$ is the differential transition rate.
The nonequilibrium distribution function for the i-th species $F_i$ is given by
\begin{equation}
F_i = f_i - \phi_i \frac{\partial f_i} {\partial \epsilon_i}~,
\end{equation}
where the equilibrium Fermi-Dirac distribution function  
$f(\epsilon_i)~=~\frac 1{1~+~e^{\frac{\epsilon_i~-~\mu_i}{kT}}}$ and the
departure from the equilibrium is given by $\phi$. We adopt the following 
ansatz for $\phi_i$ \cite{Yak,Rup}
 
\begin{equation}
\phi_i = -\tau_i (v_{i}p_{j} - \frac{1}{3} v_i p_i \delta_{ij})
(\bigtriangledown_{i} {\cal V}_j + \bigtriangledown_{j} {\cal V}_i - 
\frac{2}{3}\delta_{ij} {\vec {\bigtriangledown}} \cdot {\vec {\cal V}})~,
\end{equation}
where $\tau_i$ is the effective relaxation time for the ith species and 
${\cal V}$
is the flow velocity. The transport equations are linearised and multiplied by
$(2\pi\hbar)^{-3} (v_{i}p_{j} - \frac{1}{3} v_i p_i \delta_{ij}) d{\bf p}_i$. 
Summing over spin $s_i$ and integrating over $d{\bf p}_i$ 
we obtain a set of relations between effective relaxation times and collision
frequencies \cite{Yak}
\begin{equation}
\sum_{j=n,p,e,\mu} (\nu_{ij}\tau_i + \nu'_{ij}\tau_j) = 1~,
\end{equation}
and
the effective collision frequencies are
\begin{eqnarray}
\nu_{ij} = \frac{3\pi^2\hbar^3}{{2p_{F_i}^5{kT}m_i^*}(1+\delta_{ij})}
{\sum_{{s_i},{s_{i'}},{s_j},{s_{j'}}}}
\int \frac {d{\bf p}_id{\bf p}_jd{\bf p}_{i'}d{\bf p}_{j'}}{(2\pi\hbar)^{12}}
W_{ij} [f_if_j(1-f_{i'})(1-f_{j'})] \nonumber\\ 
\times [\frac{2}{3}p_i^4 + \frac{1}{3}p_i^2p_{i'}^2 
- ({\bf p}_i \cdot {\bf p}_{i'})^2]~,
\end{eqnarray}
\begin{eqnarray}
\nu'_{ij} = \frac{3\pi^2\hbar^3}{{2p_{F_i}^5{kT}m_j^*}(1+\delta_{ij})}
{\sum_{{s_i},{s_{i'}},{s_j},{s_{j'}}}}
\int \frac {d{\bf p}_id{\bf p}_jd{\bf p}_{i'}d{\bf p}_{j'}}{(2\pi\hbar)^{12}}
W_{ij} [f_if_j(1-f_{i'})(1-f_{j'})]\nonumber\\ 
\times [\frac{1}{3}p_i^2p_{j'}^2 - \frac{1}{3}p_i^2p_j^2 
+ ({\bf p}_i \cdot {\bf p}_j)^2 - ({\bf p}_i \cdot {\bf p}_{j'})^2]~.
\end{eqnarray}
The differential transition rate is given by 
\begin{equation}
{\sum_{{s_i},{s_{i'}},{s_j},{s_{j'}}}} W_{ij} =  
{4(2\pi)^4\hbar^2} 
\delta(\epsilon_i+\epsilon_j-\epsilon_{i'} - \epsilon_{j'})
\delta({\bf p}_i+{\bf p}_j - {\bf p}_{i'} - {\bf p}_{j'})
{\cal Q}_{ij}~, 
\end{equation}
where ${\cal Q}_{ij}=<|{\cal M}_{ij}|^2>$ is the squared matrix element summed
over final spins and averaged over initial spins \cite{Yak,Yak2,Bai}.

We obtain effective relaxation times for different particle species solving a
matrix equation that follows from Eq.(6). The matrix equation has 
the following form:
\begin{equation}
\left(\begin{array}{llll}
\nu_e & \nu'_{e\mu} & \nu'_{ep} & 0\\
\nu'_{{\mu}e} & \nu_{{\mu}} & \nu'_{{\mu}p} & 0\\
\nu'_{pe} & \nu'_{p{\mu}} & \nu_p & \nu'_{pn}\\
0 & 0 & \nu'_{np} & \nu_n
\end{array} \right)
\left(\begin{array}{c}
\tau_e\\
\tau_{\mu}\\
\tau_p\\
\tau_n
\end{array} \right) = 1
\end{equation} 
where,
\begin{eqnarray}
\nu_e &=& \nu_{ee} + \nu'_{ee} + \nu_{e\mu} + \nu_{ep}~,\\
\nu_{\mu} &=& \nu_{\mu\mu} + \nu'_{\mu\mu} + \nu_{{\mu}e} + \nu_{{\mu}p}~,\\
\nu_p &=& \nu_{pp} + \nu'_{pp} + \nu_{pn} + \nu_{pe} + \nu_{p{\mu}}~,\\
\nu_n &=& \nu_{nn} + \nu'_{nn} + \nu_{np}~.
\end{eqnarray}
It is to be noted here that the proton-proton interaction is made up of 
contributions from electromagnetic and strong interactions. As there is no 
interference of the electromagnetic and strong interaction terms, 
the differential transition rate for the proton-proton scattering is the sum of 
electromagnetic and strong contributions. This was discussed earlier in 
Ref.\cite{Flo2}. Therefore, we can write the strong and electromagnetic parts 
of the effective collision frequencies of proton-proton scattering as
\begin{eqnarray}
\nu_{pp}& = & \nu_{pp}^{s} + \nu_{pp}^{em}~, \nonumber\\
\nu{'}_{pp}& = & \nu{'}_{pp}^{s} + \nu{'}_{pp}^{em}~.
\end{eqnarray}
Here the superscripts '$em$' and '$s$' denote the electromagnetic and strong 
interactions. Solutions of Eq. (10) are given below
\begin{eqnarray}
\tau_e = \frac{(\nu_p\nu_n -\nu'_{pn} \nu'_{np}) 
(\nu_{\mu}-\nu'_{e{\mu}}) + (\nu'_{pn} - \nu_n) (\nu_{\mu}\nu'_{ep}-\nu'_{e\mu}
\nu'_{{\mu}p}) + \nu_n \nu'_{p\mu} (\nu'_{ep} - \nu'_{{\mu}p})} {detA}
~,\nonumber\\
\tau_{p} = \frac{(\nu_n - \nu'_{pn}) (\nu_e \nu_{\mu} - \nu'_{e\mu}
\nu'_{{\mu}e}) + \nu'_{p\mu} \nu_n (\nu'_{e{\mu}} - \nu_e) + \nu'_{pe} \nu_n 
(\nu'_{e{\mu}} - \nu_{{\mu}})}{det A} 
~,\nonumber\\
\tau_{n} = \frac{(\nu_p - \nu'_{np}) (\nu_e \nu_{\mu} - \nu'_{e\mu}
\nu'_{{\mu}e})
+ (\nu'_{np} - \nu'_{{\mu}p}) (\nu'_{p{\mu}} \nu_e - \nu'_{e\mu}\nu'_{p{\mu}}) 
+ (\nu'_{ep} - \nu'_{np}) 
(\nu'_{{\mu}e} \nu'_{p\mu} - \nu'_{pe} \nu_{{\mu}})}{det A} ~.
\end{eqnarray}
where $A$ is the $4\times4$ matrix of Eq. (10) and $det A = [\nu_e \nu_{\mu} 
(\nu_p \nu_n - \nu'_{pn}\nu'_{np}) - \nu_e 
\nu'_{{\mu}p} \nu'_{p\mu} \nu_n - \nu'_{e\mu} \nu'_{{\mu}e} 
(\nu_p \nu_n - \nu'_{pn} \nu'_{np}) - \nu'_{e\mu} \nu'_{{\mu}p} \nu'_{pe} \nu_n
+ \nu'_{ep} \nu'_{{\mu}e} \nu'_{p\mu} \nu_n - \nu'_{ep} \nu_{\mu} \nu'_{pe} 
\nu_n$]. We obtain $\tau_{\mu}$ from $\tau_e$ replacing $e$ by $\mu$.
In the next paragraphs, we discuss the determination of matrix element squared
for electromagnetic and strong interactions. 

First we focus on the electromagnetic scattering of charged particles. Here
we adopt the plasma screening of the interaction due to the exchange of
longitudinal and transverse plasmons as described in Refs.\cite{Yak,Yak2,Hei}.
The matrix element for the collision of identical charged particles is given by
$M_{12} = M^{(1)}_{12} + M^{(2)}_{12}$, where the first term implies the
scattering channel $1 2 \rightarrow 1' 2'$ and the second term corresponds to 
that of $1 2 \rightarrow 2' 1'$. The scattering of charged particles in 
neutron star interiors involves 
small momentum and energy transfers.
Consequently both channels contribute equally because the interference term is
small in this case. The matrix element for nonidentical particles is given by
\cite{Yak,Yak2,Hei}
\begin{equation}
M_{12 \rightarrow 1'2'} = \frac{4\pi e^2}{c^2} \left(\frac{J_{0}^{11'} 
J_{0}^{22'}} {q^2 + \Pi_l^2} -
\frac{{\bf J}_{t}^{11'} \cdot {\bf J}_{t}^{22'}} 
{q^2 - \omega^2/c^2 + \Pi_t^2} \right)~,
\end{equation}
where $\bf q$ and $\omega$ are momentum and energy transfers in the neutron 
star interior. Further 
four-current 
$(J_0, {\bf J})$ and longitudinal and transverse polarization
functions $(\Pi_l, \Pi_t)$ have the same expressions as defined in 
Ref.\cite{Yak}. 
It is to be noted that ${\bf J}_t$ is the transverse component of $\bf J$ with
respect to $\bf q$ and the longitudinal component is related to the timelike
component $J_0$ by the conservation of current \cite{Hei}. 
Polarization functions $\Pi_l$ and $\Pi_t$ are associated with the plasma 
screening 
of charged particles' interactions through the exchange of longitudinal and 
transverse plasmons, respectively. After evaluating the matrix element squared
and doing the angular and energy integrations, the effective collision 
frequencies are calculated following the prescription of Ref.\cite{Yak,Yak2}.
The collision frequencies of Eqs. (7) and (8) for charged particles become
\begin{eqnarray}
\nu_{ij} &=& \nu_{ij}^{||} + \nu_{ij}^{\perp}~,\nonumber\\
\nu'_{ij} &=& {\nu'}_{ij}^{||\perp}~,
\end{eqnarray}
where $\nu_{ij}^{||}$ and $\nu_{ij}^{\perp}$ correspond to the charged particle
interaction due to the exchange of longitudinal and transverse plasmons and 
$\nu_{ij}^{||\perp}$ is the result of the interference of both interactions. 
For small momentum and energy transfers,  different components of the collision
frequency are given by \cite{Yak,Yak2} 
\begin{eqnarray}
\nu_{ij}^{\perp} &=& \frac{e^4 \alpha}{\hbar^4 c^3} \frac{p_{F_j}^2}
{p_{F_i}m_i^* c} \left({\frac{\hbar c}{q_t^2}}\right)^{1/3} (kT)^{5/3},
\nonumber\\
\nu_{ij}^{||} &=& \frac{e^4 \pi^2 m_i^* m_j^{*2}} 
{\hbar^4 p_{F_i}^3 q_l} (kT)^2~, \nonumber\\
\nu{'}_{ij}^{||\perp} &=& \frac{2 e^4 \pi^2 m_i^* p_{F_j}^2} 
{\hbar^4 c^2 p_{F_i}^3 q_l} (kT)^2~,
\end{eqnarray}
where $i,j=e,\mu,p$ and longitudinal and transverse wave numbers are given by
\begin{eqnarray}
q_l^2 &=& \frac{4e^2}{\hbar^3 c \pi} \sum_{j=e,\mu,p}{c m_j^* p_{F_j}}~,
\nonumber\\
q_t^2 &=& \frac{4e^2}{\hbar^3 c \pi} \sum_{j=e,\mu,p}p_{F_j}^2~.
\end{eqnarray} 
The value of $\alpha = 2 ({\frac{4}{\pi}})^{1/3} \Gamma (8/3) \zeta (5/3) 
\sim 6.93$ where $\Gamma (x)$ and $\zeta (x)$ are gamma and Riemann zeta 
functions, respectively. The shear viscosities of electrons and muons 
are given by 
\cite{Yak}
\begin{equation}
\eta_{i(=e,\mu)} = \frac {n_i p_{F_i}^2 \tau_i}{5m_i^*}~.
\end{equation}
Here effective masses ($m_i^*$) of electrons and muons are equal to their
corresponding chemical potentials because of relativistic effects.
It was noted
that the shear viscosity was reduced due to the inclusion of plasma screening
by the exchange of transverse plasmons \cite{Yak,Yak2}. It is worth mentioning 
here that we extend the calculation of the collision frequencies 
for electrons and muons in Refs.\cite{Yak,Yak2} to that of protons due to 
electromagnetic interaction. Before the appearance of the condensate in our 
calculation, protons may be treated as passive scatterers as was earlier done 
by Ref.\cite{Yak}. However, after the onset of the antikaon condensation, 
electrons and muons are replaced by $K^-$ mesons and proton fraction increases 
rapidly in the system \cite{Pal,Banik}. In this situation protons can not be 
treated as passive scatterers. 
  
Next we focus on the calculation of collision frequencies of neutron-neutron, 
proton-proton and neutron-proton scatterings due to the strong interaction. The
knowledge of nucleon-nucleon scattering cross sections are exploited in this
calculation. This was first done by Ref.\cite{Flo2}. Later recent developments
in the calculation of nucleon-nucleon scattering cross sections in the 
Dirac-Brueckner approach were considered for this purpose \cite{Yak2,Bai}. Here
we adopt the same prescription of Ref.\cite{Bai} for the calculation of 
collision frequencies due to nucleon-nucleon scatterings. The collision 
frequency for the scattering of identical particles under strong interaction is
given by
\begin{equation}
\nu_{ii} + \nu{'}_{ii} = \frac{16 m_i^{*3} (kT)^2}{3m_n^2 {\hbar}^3} 
S_{ii}~, 
\end{equation}     
\begin{equation}
S_{ii} = \frac{m_n^2}{16{\hbar}^4{\pi}^2} \int_{0}^{1} dx'
\int_{0}^{\sqrt{(1-x'^2)}} dx
\frac{12{x^2}{x'^2}}{\sqrt{1-x^2 - x'^2}} {\cal{Q}}_{ii}~,
\end{equation}
where $i=n,p$ and $m_n$ is the bare nucleon mass and 
${\cal Q}_{ii}$ is the matrix element squared which appears in Eq. (9).
Similarly we can write the collision frequency for nonidentical particles as
\begin{eqnarray}
\nu_{ij} = \frac{32 m_i^{*} m_j^{*2} (kT)^2}{3m_n^2 {\hbar}^3} S_{ij}~,
\nonumber\\ 
\nu{'}_{ij} = \frac{32 m_i^{*2} m_j^{*} (kT)^2}{3m_n^2 {\hbar}^3} S'_{ij}~, 
\end{eqnarray}     
and
\begin{eqnarray}
S_{ij} = \frac{m_n^{2}}{16{\hbar}^4{\pi}^2} \int_{0.5-x_0}^{0.5+x_0}dx' 
\int_{0}^{f} dx
\frac{6(x^2 - x^4)}{\sqrt{(f^2 - x^2)}} {\cal{Q}}_{ij}~,\nonumber\\
S'_{ij} = \frac{m_n^{2}}{16{\hbar}^4{\pi}^2} \int_{0.5-x_0}^{0.5+x_0}dx' 
\int_{0}^{f} dx
\frac{[6{x^4}+ 12 {x^2}{x'^2} - (3 + 12x_0^2)x^2]}{\sqrt{(f^2 - x^2)}}
{\cal{Q}}_{ij}~.
\end{eqnarray}
We define $x_0 = \frac{p_{F_j}}{2p_{F_i}}$, $x=\frac{{\hbar}q}{2p_{F_i}}$,
$x'=\frac{{\hbar}q'}{2p_{F_i}}$, $f = \frac{\sqrt{x_0^2 - (0.25 + x_0^2 - 
x'^2)^2}}{x'}$, where momentum transfers 
${\bf q} = {\bf p}_{j'} - {\bf p}_{j}$ and 
${\bf q}' = {\bf p}_{j'} - {\bf p}_{i}$. We find that the calculation of 
$S_{ij}$,
$S_{ii}$ and $S'_{ij}$ requires the knowledge of ${\cal Q}_{ii}$ and 
${\cal Q}_{ij}$. The
matrix elements squared may be extracted from nucleon-nucleon differential
cross sections. A detailed discussion on the calculation of matrix elements 
squared from the in-vacuum nucleon-nucleon differential scattering cross 
sections computed using Dirac-Brueckner approach \cite{Mach1,Mach2} can be found
in Ref.\cite{Yak2,Bai}. We follow this procedure in this calculation. It is to 
be noted here that $S_{pp}$, $S_{pn}$ and $S'_{ij}$ are the new results of 
this calculation. As soon as we know the collision frequencies of 
nucleon-nucleon scatterings due to the strong interaction, we can immediately
calculate effective relaxation times of neutrons and protons from Eq. (16). 
This leads to the calculation of the neutron and proton shear viscosities as  
\begin{eqnarray}
\eta_{n} = \frac {n_n p_{F_n}^2 \tau_n}{5m_n^*}~,\nonumber\\
\eta_{p} = \frac {n_p p_{F_p}^2 \tau_p}{5m_p^*}~.
\end{eqnarray}
Finally the total shear viscosity is given by
\begin{equation}
\eta_{total} = \eta_n + \eta_p + \eta_e + \eta_{\mu}~.
\end{equation}

The EOS enters into the calculation of the shear viscosity as
an input. We construct the EOS within the framework of the relativistic field 
theoretical model \cite{walecka, serot}. Here we consider a first order phase
transition from nuclear matter to $K^-$ condensed matter. We adopt the Maxwell
construction for the first order phase transition. The constituents of matter
are neutrons, protons, electrons and muons in both phases and 
also (anti)kaons in the $K^-$ condensed phase. Both phases maintain charge
neutrality and $\beta$ equilibrium conditions. 
Baryons and (anti)kaons are interacting with each other and among
themselves by the exchange of $\sigma$, $\omega$ and $\rho$ 
mesons \cite{Pal,Banik}. The baryon-baryon interaction is given by the
Lagrangian density \cite{glendenning,schaffnerprc}
\begin{eqnarray}
{\cal L}_B &=& \sum_{B=n,p} \bar\psi_{B}\left(i\gamma_\mu 
{\partial}^{\mu} - m_B
+ g_{\sigma B} \sigma - g_{\omega B} \gamma_\mu \omega^\mu
- g_{\rho B}
\gamma_\mu{\mbox{\boldmath t}}_B \cdot
{\mbox{\boldmath $\rho$}}^\mu \right)\psi_B\nonumber\\
&& + \frac{1}{2}\left( \partial_\mu \sigma\partial^\mu \sigma
- m_\sigma^2 \sigma^2\right) - U(\sigma) \nonumber\\
&& -\frac{1}{4} \omega_{\mu\nu}\omega^{\mu\nu}
+\frac{1}{2}m_\omega^2 \omega_\mu \omega^\mu
- \frac{1}{4}{\mbox {\boldmath $\rho$}}_{\mu\nu} \cdot
{\mbox {\boldmath $\rho$}}^{\mu\nu}
+ \frac{1}{2}m_\rho^2 {\mbox {\boldmath $\rho$}}_\mu \cdot
{\mbox {\boldmath $\rho$}}^\mu ~.
\end{eqnarray}
The scalar self-interaction \cite{schaffnerprc,glendenning,boguta} is 
\begin{equation}
U(\sigma)~=~\frac13~g_1~m_N~(g_{\sigma N}\sigma)^3~+~ \frac14~g_2~
(g_{\sigma N}\sigma)^4~,
\end{equation}
The effective nucleon mass is given by $m_B^* = m_B - g_{\sigma B} \sigma$, 
where $m_B$ is the vacuum baryon mass.
The Lagrangian density for (anti)kaons in the minimal coupling is given by
\cite{Pal,Banik,Gle99}
\begin{equation}
{\cal L}_K = D^*_\mu{\bar K} D^\mu K - m_K^{* 2} {\bar K} K ~,
\end{equation}
where the covariant derivative is
$D_\mu = \partial_\mu + ig_{\omega K}{\omega_\mu}
+ i g_{\rho K}
{\mbox{\boldmath t}}_K \cdot {\mbox{\boldmath $\rho$}}_\mu$ and
the effective mass of (anti)kaons is
$m_K^* = m_K - g_{\sigma K} \sigma$.
The in-medium energies of $K^{\pm}$ mesons are given by
\begin{equation}
\omega_{K^{\pm}} =  \sqrt {(p^2 + m_K^{*2})} \pm \left( g_{\omega K} \omega_0
+ \frac {1}{2} g_{\rho K} \rho_{03} \right)~.
\end{equation}
The condensation sets in when the chemical potential of $K^-$ mesons  
($\mu_{K^-} = \omega_{K^-}$) is equal to the electron chemical potential i.e. 
$\mu_e = \mu_{K^-}$.

Using the mean field approximation \cite{walecka,serot} and solving 
equations of motion self-consistently,
we calculate the effective nucleon mass and Fermi momenta of particles
at different baryon densities.

\section{Results and Discussions}
The knowledge of meson-nucleon and meson-kaon coupling constants are needed for 
this calculation. The nucleon-meson coupling constants determined by 
reproducing the nuclear matter saturation
properties such as binding energy $E/B=-16.3$ MeV, baryon density $n_0=0.153$ 
fm$^{-3}$, the asymmetry energy coefficient $a_{\rm asy}=32.5$ MeV, 
incompressibility $K=300$ MeV ,and effective nucleon mass $m^*_N/m_N = 0.70$, 
are taken from Ref.\cite{Gle91}. Next
we determine the kaon-meson coupling constants using
the quark model and isospin counting rule. The vector coupling constants are
given by
\begin{equation}
g_{\omega K} = \frac{1}{3} g_{\omega N} ~~~~~ {\rm and} ~~~~~
g_{\rho K} = g_{\rho N} ~.
\end{equation}
The scalar coupling constant is obtained from the real part of
$K^-$ optical potential depth at normal nuclear matter density
\begin{equation}
U_{\bar K} \left(n_0\right) = - g_{\sigma K}\sigma - g_{\omega K}\omega_0 ~.
\end{equation}
It is known that antikaons experience an 
attractive potential and kaons have a repulsive interaction in nuclear matter 
\cite{Fri94,Fri99,Koc,Waa,Li,Pal2}. On the one hand, the analysis of $K^-$ 
atomic data indicated that
the real part of the antikaon optical potential could be as large as 
$U_{\bar K} = -180 \pm 20$ MeV at normal nuclear matter density
\cite{Fri94,Fri99}. On the other hand, chirally motivated coupled channel 
models with a self-consistency requirement predicted shallow potential depths 
of $-40$-$60$ MeV \cite{Ram,Koch}. Recently, the double pole structure of 
$\Lambda (1405)$ was investigated in connection with the antikaon-nucleon
interaction \cite{Mag,Hyo}. Further, the
highly attractive potential depth of several hundred MeV was obtained in the 
calculation of deeply bound antikaon-nuclear states \cite{Yam,Akai}. An 
alternative explanation to the deeply bound antikaon-nuclear states was given
by others \cite{Toki}. This shows that the value of antikaon optical potential 
depth is still a debatable issue. Motivated by the findings of the analysis of 
$K^-$ atomic data, we perform this calculation for an antikaon optical 
potential depth $U_{\bar K} = -160$ MeV at normal nuclear matter density. We 
obtain kaon-scalar meson coupling constant $g_{\sigma K} = 2.9937$ 
corresponding to $U_{\bar K}(n_0) = -160$ MeV. 

The composition of neutron star matter including the $K^-$ condensate as a 
function of normalised baryon density is shown in Fig. 1. The $K^-$ 
condensation sets in at 2.43$n_0$. Before the onset of the condensation, all
particle fractions increase with baryon density. In this case, the charge 
neutrality is maintained by protons, electrons and muons. As soon as the 
antikaon condensate is formed, the density of $K^-$ mesons in the condensate 
rapidly increases and $K^-$ mesons replace leptons in the system. The proton 
density eventually becomes equal to the $K^-$ density. The proton density in 
the
presence of the condensate increases significantly and may be higher than the
neutron density at higher baryon densities \cite{Pal2}. This increase in the
proton fraction in the presence of the $K^-$ condensate might result in an 
enhancement 
in the proton shear viscosity and appreciable reduction in the electron and 
muon viscosities compared with the case without the condensate. We discuss this
in details in the following paragraphs.

Next we focus on the calculation of $\nu_{ii}$, $\nu_{ij}$ and $\nu'_{ij}$. For
the scatterings via the electromagnetic interaction, we calculate those 
quantities using Eqs. (18) and (19). On the other hand, $\nu$s corresponding 
to collisions through the strong interaction are estimated using Eqs. 
(22)-(25). In an earlier calculation, the authors considered
only $S_{nn}$ and $S_{np}$ \cite{Yak} for the calculation of the neutron shear
viscosity in nucleons-only neutron star matter because protons were treated as
passive scatterers. It follows from the discussion in the preceding paragraph
that protons can no longer be treated as passive scatterers because of the 
large proton fraction in the presence of the $K^-$ condensate. Consequently the
contributions of $S_{pp}$ and $S_{pn}$  have to be taken into account in the
calculation of the proton and neutron shear viscosities. The expressions of
$S_{nn}$, $S_{pp}$, $S_{np}$ and $S_{pn}$ given by Eqs. (23) and (25) involve
matrix elements squared. We note that there is an one to one correspondence 
between the differential cross section and the matrix element squared 
\cite{Bai}. We exploit the in-vacuum nucleon-nucleon cross sections of Li and
Machleidt \cite{Mach1,Mach2} calculated using Bonn
interaction in the Dirac-Brueckner approach for the calculation of matrix 
elements squared. We fit the neutron-proton as well as proton-proton 
differential cross sections and use them in Eqs. (23) and (25) to
calculate $S_{nn}$, $S_{pp}$, $S_{np}$ and $S_{pn}$ which are functions of 
neutron ($p_{F_n}$) and proton ($p_{F_p}$) Fermi momenta. The values of 
$p_{F_n}$ ranges from 1.3 to 2.03 $fm^{-1}$ whereas that of $p_{F_p}$ spans the
interval 0.35 to 1.73 $fm^{-1}$. This corresponds to the density range 
$\sim$0.5 to $\sim$ 3.0$n_0$. We fit the results of our calculation. 
Figures 2 and 3 display the variation of $S_{nn}$, $S_{pp}$, $S_{np}$ and 
$S_{pn}$ with baryon density. The value of $S_{nn}$ is greater than that of
$S_{np}$ in the absence of the condensate as evident from Fig. 2. Our results
agree well with those of Ref.\cite{Yak}. However, $S_{np}$ rises rapidly with
baryon density after the onset of the $K^-$ condensation and becomes higher 
than $S_{nn}$. It is noted that the effect of the condensate on $S_{nn}$ is not
significant. Figure 3 shows that $S_{pp}$ drops sharply with increasing baryon
density and crosses the curve of $S_{pn}$ in the absence of the condensate.
However $S_{pp}$ and $S_{pn}$ are not influenced by the antikaon condensate. A 
comparison of Fig. 2 and Fig. 3 reveals that $S_{pp}$ is almost
one order of magnitude larger than $S_{nn}$ at lower baryon densities. This 
may be attributed to the smaller
proton Fermi momentum. We also compute $S'_{np}$ and $S'_{pn}$ (not shown here)
and these quantities have negative values. Further we find that the magnitude 
of $S'_{pn}$ is higher than that of $S'_{np}$. It is to be noted here that 
$S'_{ij}$ is related to ${\nu}'_{ij}$ by Eq. (24). This is again connected to 
Eq. (6). Therefore, the values of $S'_{ij}$ and ${\nu}'_{ij}$ can be made 
positive by putting a negative sign between two terms in Eq. (6). 

As soon as we know $\nu$s, we can calculate effective relaxation times using 
Eq. (16) and shear viscosities using Eqs. (21), (26) and (27). First, 
we discuss the total shear viscosity in nuclear matter without a $K^-$ 
condensate. This is shown as a function of baryon density at a temperature 
$10^8$K in Fig. 4. Here our results indicated by the solid line are compared
with the calculation of the total shear viscosity using the EOS of Akmal, 
Pandharipande and Ravenhall (APR) \cite{APR} denoted by the dotted line and 
also with the 
results of Flowers and Itoh \cite{Flo1,Flo2}. For the APR 
case, we exploit the parametrization of the EOS by Heiselberg and 
Hjorth-Jensen \cite{HHJ}. Further we take density independent nucleon effective
masses $m_n^* = m_p^* = 0.8 m_n$ for the calculation with the APR EOS which was
earlier discussed by Shternin and Yakovlev \cite{Yak2}. On the other hand, the 
results 
of Flowers and Itoh were parametrized by Cutler and Lindblom (CL) \cite{Cut}
and it is shown by the dashed line in Fig. 4. It is evident from Fig. 4 that 
the total shear viscosity in our calculation is significantly higher than other
cases. This may be attributed to the fact that our EOS is a fully
relativistic one. 
   
We exhibit shear viscosities in the presence of an antikaon condensate as a 
function of baryon density in Fig. 5. This calculation is
performed at a temperature $10^8$K. In the absence of the
$K^-$ condensate, the contribution of the electron shear viscosity to the 
total shear viscosity is the highest. The electron, muon and neutron shear 
viscosities exceed the proton shear viscosity by several orders of magnitude.
Further we note that the lepton viscosities are greater than
the neutron viscosity. On the other hand, we find interesting results in the
presence of the antikaon condensate. The electron and muon shear
viscosities decrease very fast after the onset of $K^-$ condensation whereas 
the proton shear viscosity rises in this case. There is almost no change in 
the neutron shear 
viscosity. It is interesting to note that the proton shear viscosity in the
presence of the condensate approaches the value of the neutron shear viscosity
as baryon density increases. The total shear viscosity decreases in the $K^-$
condensed matter due to the sharp drop in the lepton shear viscosities. Here
the variation of shear viscosities with baryon density is shown up to 3$n_0$. 
The neutron and proton shear viscosities in neutron star matter with the $K^-$ 
condensate might dominate over the electron and muon shear viscosities beyond 
baryon density 3$n_0$. Consequently, the total shear viscosity would again 
increase.

The temperature dependence of the total shear viscosity is shown in Fig. 6. 
In an earlier calculation, electron and muon shear viscosities were determined
by collisions only due to the exchange of transverse plasmons because
this was the dominant contribution \cite{Yak2}. Under 
this approximation, the electron and muon shear viscosities had a temperature 
dependence of $T^{-5/3}$, whereas, the neutron shear viscosity was 
proportional to
$T^{-2}$. The temperature dependence of the electron and muon 
shear viscosities deviated from the standard temperature dependence of 
the shear viscosity of neutron Fermi liquid. However, in this calculation we 
have not made any such approximation. We have considered all the components of 
effective collision frequency which have different temperature dependence as 
given by Eq. (19). This gives rise to a complicated temperature dependence in 
the calculation of shear viscosity.  The total shear viscosity is plotted for 
$T = 10^7$, $10^8$, and $10^9$ K in Fig. 6. It is noted that the shear 
viscosity increases as temperature decreases.

The shear viscosity plays an important role in damping the r-mode instability
in old and accreting neutron stars \cite{Nay,Chat1,Chat2,Chat3,Chat4}. The 
suppression of the instability is achieved by the
competition of various time scales associated with gravitational radiation
($\tau_{GR}$), hyperon bulk viscosity ($\tau_{B}$), modified Urca bulk 
viscosity ($\tau_{U}$), and shear viscosity ($\tau_{SV}$). At high temperatures
the bulk viscosity damp the r-mode instability. As neutron stars cool down, the
bulk viscosity might not be the dominant damping mechanism. The shear viscosity
becomes significant in the temperature regime $\leq 10^{8}$K and might suppress
the r-mode instability effectively. 

In this calculation, we consider only the antikaon optical potential depth
$U_{\bar K} = -160$ MeV. However, this calculation could be performed for
other values of antikaon optical potential depths. As the magnitude of the
$K^-$ potential depth decreases, the threshold of the antikaon condensation is
shifted to higher densities \cite{Pal}. On the other hand, hyperons may also
appear in neutron star matter around 2-3$n_0$. Negatively charged hyperons 
might delay the onset of $K^-$ condensation \cite{schaffnerprc,Ell,Kno}. 
However, it was 
noted in an earlier calculation that $\Sigma^-$ hyperons were excluded from 
the system because of repulsive $\Sigma$-nuclear matter interaction and 
$\Xi^-$ hyperons might appear at very high baryon density \cite{Banik}. 
However, the appearance of $\Lambda$ hyperons could compete with the threshold 
of $K^-$ condensation. If $\Lambda$ hyperons appear before $K^-$ condensation,
the threshold of $K^-$ condensation is shifted to higher baryon density 
because of softening in the equation of state due to $\Lambda$ hyperons. But 
the qualitative results of the shear viscosity discussed above remain the 
same.  
 
\section{Summary and Conclusions}
We have investigated the shear viscosity in the presence of a $K^-$ 
condensate.
With the onset of $K^-$ condensation, electrons and muons are replaced by $K^-$
mesons rapidly. The proton fraction also increases and eventually becomes equal
to the neutron fraction in the $K^-$ condensed neutron star matter. This has 
important consequences for the electron, muon and proton shear viscosities. We
have found that the electron and muon shear viscosities drop steeply after the
formation of the $K^-$ condensate in neutron stars. On the other hand, the 
proton
shear viscosity whose contribution to the total shear viscosity was negligible 
in earlier calculations \cite{Flo2,Yak}, now becomes significant in the 
presence of the $K^-$ condensate. The proton shear viscosity would exceed the 
neutron as well as lepton shear viscosities beyond 3$n_0$. The total viscosity
would be dominated by the proton and neutron shear viscosities in this case.
This calculation may be extended to neutron stars with strong magnetic fields.
    
It is worth mentioning here that we adopt the Maxwell
construction for the first order phase transition in this calculation. Such a 
construction is 
justified if the surface tension between two phases is quite
large \cite{Alf}. Moreover the value of the surface tension between the 
nuclear and antikaon condensed phases or between the hadron and quark phases
is not known correctly. Therefore, this problem could also be studied using the
Gibbs construction \cite{NKG}. 

Besides the role of shear viscosity in damping the r-mode instability as well
as in pulsar glitches and free precession of neutron stars, it has an important
contribution in the nucleation rate of bubbles in first order phase 
transitions. It was shown earlier that the shear viscosity might control the 
initial growth rate of a bubble \cite {Kap,Bom}. This needs further study in 
connection with antikaon condensation in neutron stars. 

\section{Acknowledgments}
We thank R. Machleidt for providing us with the tables of neutron-proton and 
proton-proton differential scattering cross sections. RN and DB thank the 
Alexander von Humboldt Foundation for the support under the Research Group 
Linkage programme. We also acknowledge the warm hospitality at the 
Frankfurt Institute for Advanced Studies where a part of this work was 
completed.

\newpage

\vspace{-2.0cm}

{\centerline{
\epsfxsize=12cm
\epsfysize=14cm
\epsffile{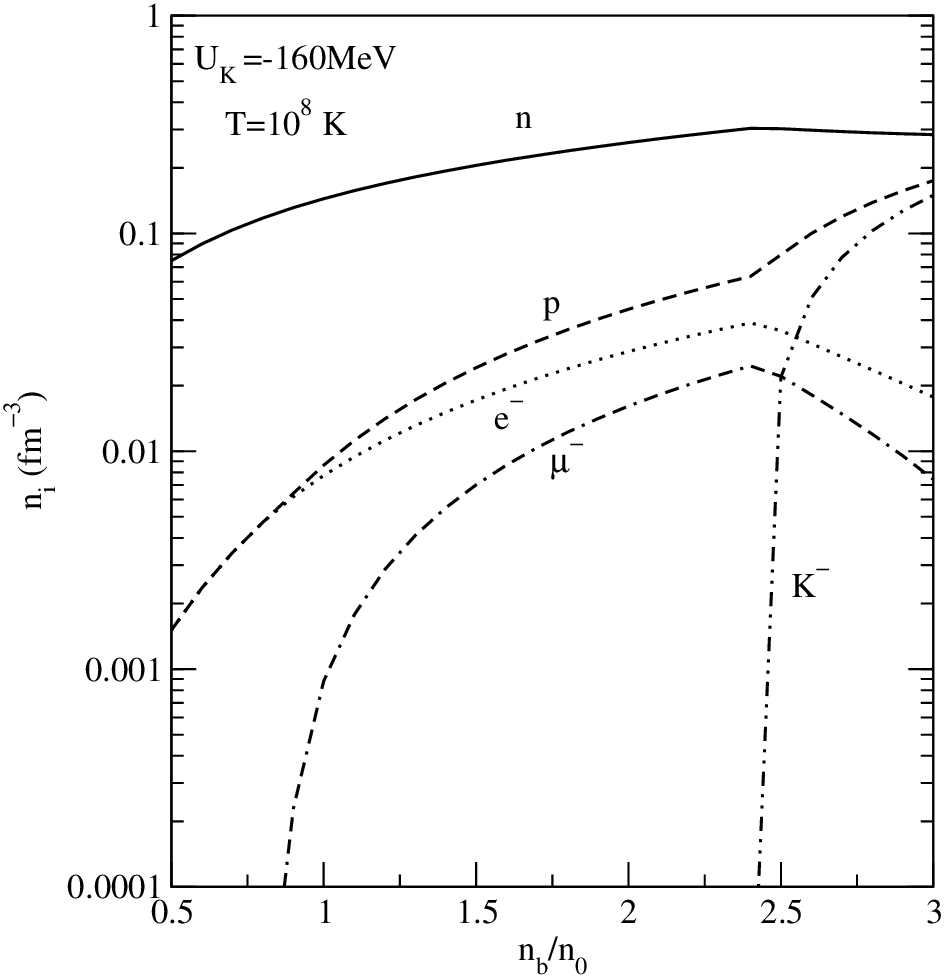}
}}

\vspace{1.0cm}

\noindent{\small{
Fig. 1. Number densities of different particle species as a function of 
normalised baryon density.}}
\label{fig:fraction}

\newpage
\vspace{-2.0cm}

{\centerline{
\epsfxsize=12cm
\epsfysize=14cm
\epsffile{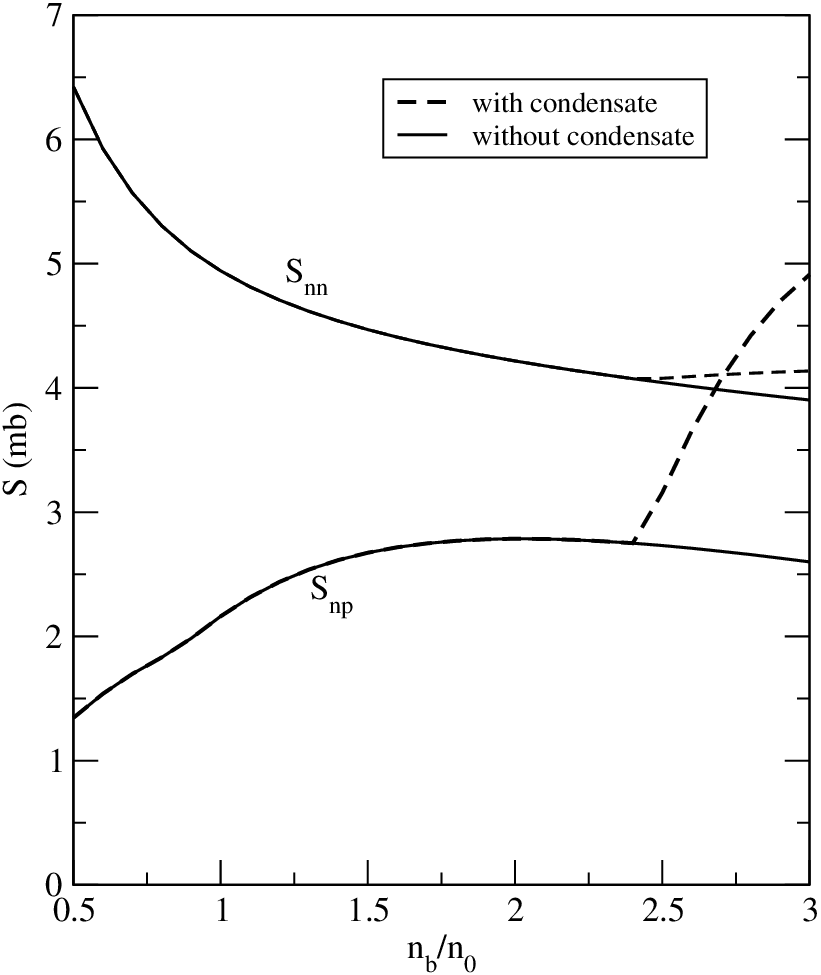}
}}

\vspace{1.0cm}

\noindent{\small{
Fig. 2. $S_{nn}$ and $S_{np}$ are plotted as a function of normalised baryon 
density.}}

\newpage
\vspace{-2.0cm}

{\centerline{
\epsfxsize=12cm
\epsfysize=14cm
\epsffile{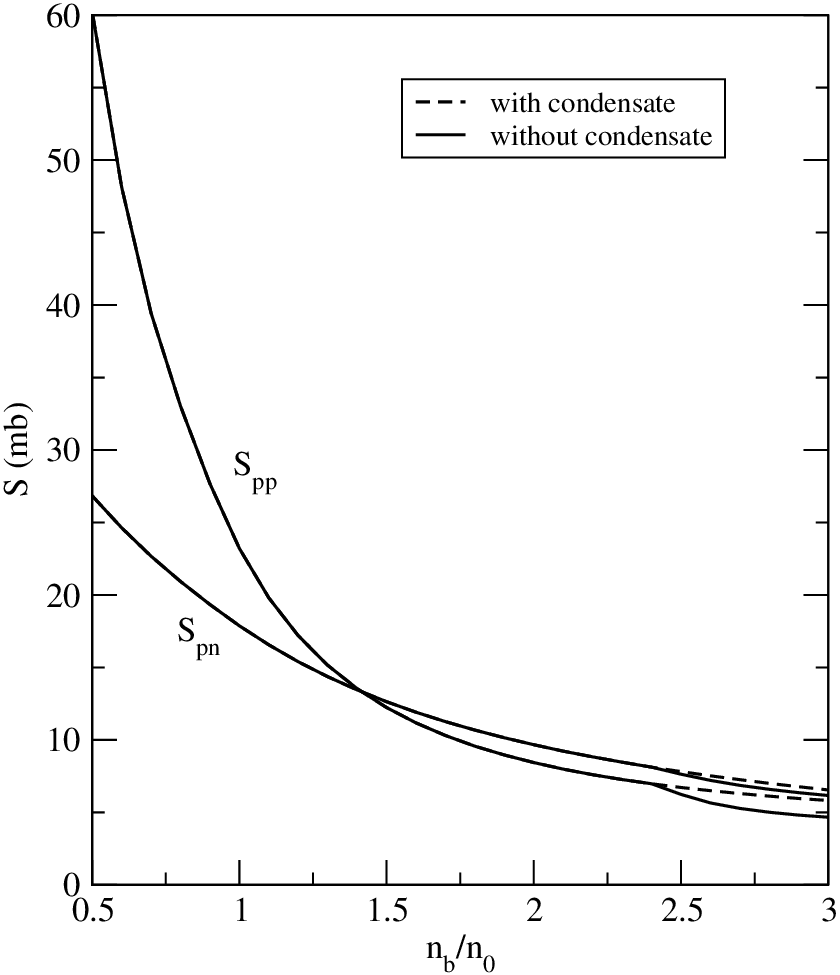}
}}

\vspace{1.0cm}

\noindent{\small{
Fig. 3. $S_{pp}$ and $S_{pn}$ are plotted as a function of normalised baryon 
density.}}

\newpage
\vspace{-2.0cm}

{\centerline{
\epsfxsize=12cm
\epsfysize=14cm
\epsffile{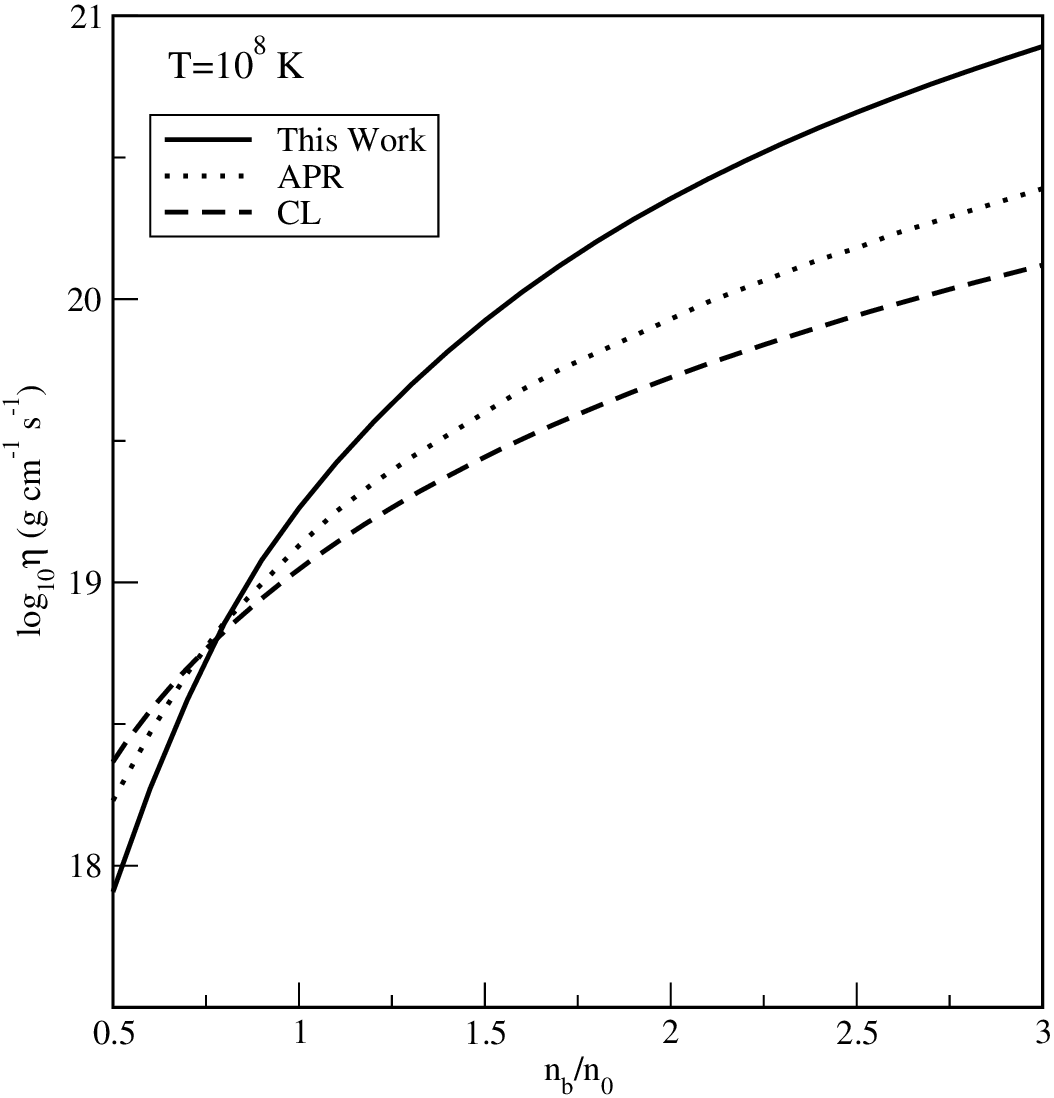}
}}

\noindent{\small{Fig. 4. Total shear viscosities in nuclear matter
without an antikaon condensate corresponding to this work (solid line), the 
parameterization
of Cutler and Lindblom (dashed line) and the EOS of Akmal, Pandharipande and
Ravenhall are shown as a function of normalised baryon density at a temperature 
$T = 10^{8}$ K.}}

\newpage

\vspace{-2.0cm}

{\centerline{
\epsfxsize=12cm
\epsfysize=14cm
\epsffile{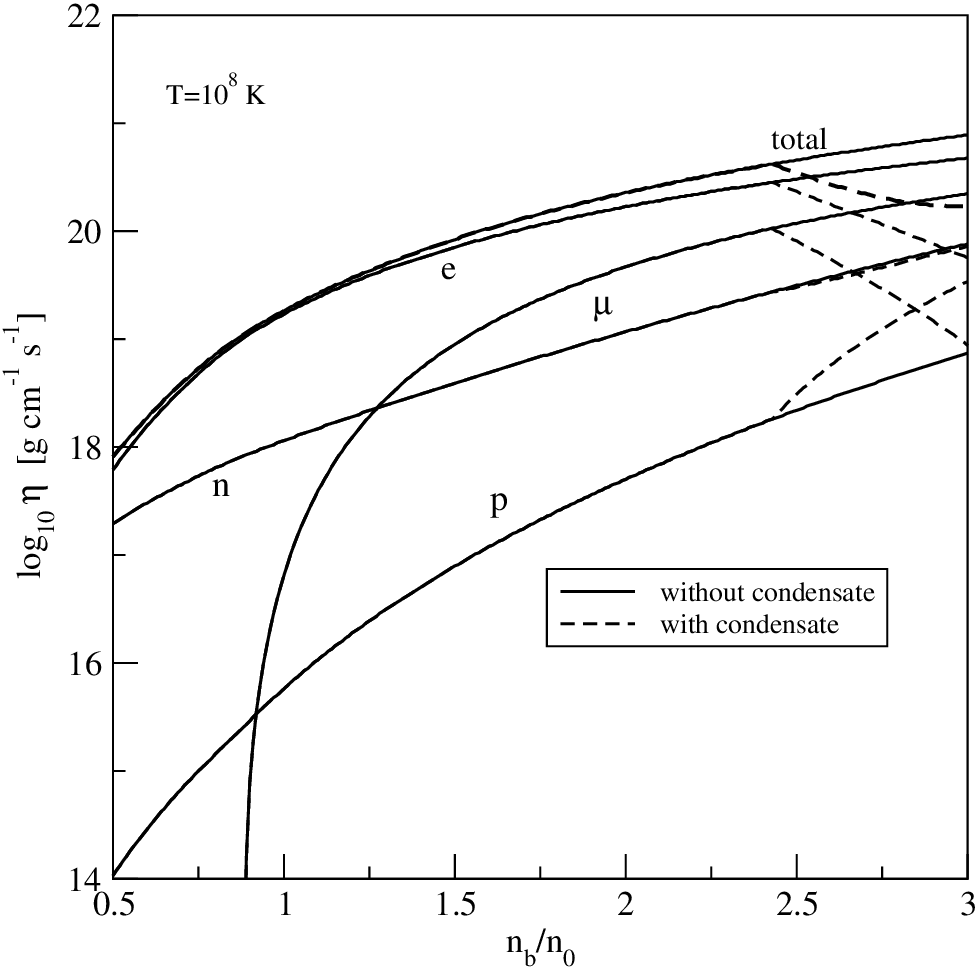}
}}

\vspace{1.0cm}

\noindent{\small{Fig. 5. The total shear viscosity as well as 
shear viscosities corresponding to different particle species
are shown as a function of normalised baryon density at a temperature 
$T = 10^{8}$ K with (solid line) and without (dashed line) a $K^-$ condensate.}}

\newpage

\vspace{-2.0cm}

{\centerline{
\epsfxsize=12cm
\epsfysize=14cm
\epsffile{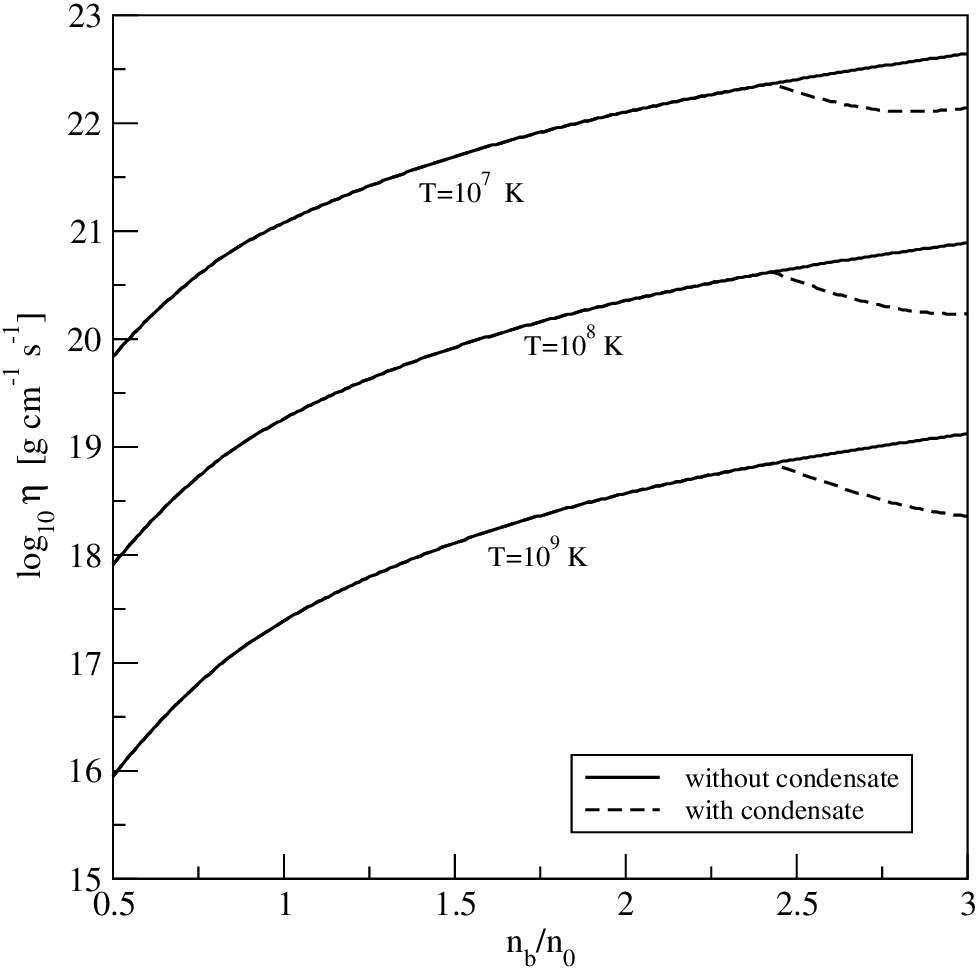}
}}

\vspace{1.0cm}

\noindent{\small{Fig. 6. The total shear viscosity 
as a function of normalised baryon density at different
temperatures with (solid line) and without (dashed line) 
a $K^-$ condensate.}}

\newpage
\end{document}